# Temperature Dependence of Spin Hall Angle of Palladium

Zhenyao Tang[1], Yuta Kitamura[1], Eiji Shikoh[2], Yuichiro Ando[1], Teruya Shinjo[1], and Masashi Shiraishi[1,*]

[1.] Graduate School of Engineering Science, Osaka University, Toyonaka 560-8531, Japan

[2.] Graduate School of Engineering, Osaka City University, Osaka 558-8585, Japan

*E-mail address: shiraishi@ee.es.osaka-u.ac.jp

**Abstract**

In this study, the temperature dependence of the spin Hall angle of palladium (Pd) was experimentally investigated by spin pumping. A $Ni_{80}Fe_{20}$/Pd bilayer thin film was prepared, and a pure spin current was dynamically injected into the Pd layer. This caused the conversion of the spin current to a charge current owing to the inverse spin Hall effect. It was found that the spin Hall angle varies as a function of temperature, whereby the value of the spin Hall angle increases to ca. 0.02 at 123 K.

In recent years, *spincurrentronics* has been attracting significant attention for potential future applications because of the low energy consumption of spin current.[1] Spin current can be generated by inducing spins into spin sink materials via spin pumping.[2–5] The spin current can then be detected by utilizing the inverse spin Hall effect (ISHE),[2,3,6] which can convert a pure spin current to a charge current by spin–orbit interaction. Many studies have applied this method in order to evaluate the spin Hall angle ($\theta_{SHE}$),[7] to inject spins into condensed matter,[8,9] and estimate the spin diffusion length and spin relaxation time,[10,11] but most of these studies have been conducted at room temperature (RT). The recent success of dynamical spin injection and transport in Si at room temperature[10] allows the establishment of spin-based logic by using a dynamical method since Si is one of the most suitable materials for beyond CMOS spin devices. In the spin logic using the dynamical method, readout of a spin current is realized by the ISHE. In the case of Si, Pt is not suitable for the "spin-charge converter", because a Pt-silicide alloy can be easily formed, which impedes the conversion. Hence, Pd is a better material for the converter. For future investigations of Si-based spin logic using the dynamical method, the temperature dependence of the logic device performance is also quite important and the temperature dependence of $\theta_{SHE}$, as a good index of spin-charge conversion, should be investigated. Herein, we report the temperature dependence of the spin Hall angle of palladium (Pd), which is a commonly used spin sink material exhibiting the ISHE. The spin Hall angle of Pd was qualitatively evaluated by changing the temperature from 130 K to RT. This evaluation yields an important spin-related physical parameter that can be investigated in future studies by applying the spin pumping method.

Figure 1(a) shows the schematic illustration of a $Ni_{80}Fe_{20}$(Py)/Pd bilayer sample. A 25-nm-thick permalloy (Py) film and a 5-nm-thick Pd film were prepared on an oxidized silicon substrate by electron beam evaporation. Both Py and Pd layers were rectangular with an area of $2 \times 1$ mm$^2$. The Pd layer was connected to the positive and negative ends of a nano-voltmeter as shown in Fig. 1. Spins from the ferromagnetic Py were injected into the Pd layer by the dynamical spin injection method. The sample system was placed at the center of a $TE_{011}$ microwave cavity in an electron spin resonance (ESR) instrument with a frequency ($f$) of 9.12 GHz. An external magnetic field $H$ was applied to the Py/Pd bilayer at an angle of $\theta_H$, as shown in Fig. 1(a). The ferromagnetic resonance (FMR) condition was determined using the following equation: [12,13]

$$\left(\frac{\omega}{\gamma}\right)^2 = H_{FMR}(H_{FMR} + 4\pi M_s), \quad (1)$$

where $\omega = 2\pi f$, $\gamma$, $H_{FMR}$, and $M_s$ are the gyromagnetic ratio of the ferromagnet, FMR field, and saturation magnetization of Py, respectively. Figure 1(b) shows the FMR spectra of Py with and without Pd. The linewidth of the spectrum corresponding to Py/Pd is larger than that corresponding to Py alone, which is attributed to the modulation of the Gilbert damping constant ($\alpha$) owing to successful spin pumping into the Pd layer. The FMR spectra and output dc voltages of the Py/Pd bilayer at RT resulting from the ISHE in the Pd are shown in Figs. 1(c) and 1(d), respectively. The FMR intensities and resonance fields were nearly identical for all values of $\theta_H$, and an electromotive force was induced when $\theta_H$ was set to 0 and 180°, while the signal was flat at $\theta_H = 90°$. Because this is consistent with the symmetry of the ISHE ($J_c \propto J_s \times \sigma$), the observed electromotive force was ascribed to the ISHE of Pd, which was due to a pure spin current generated dynamically by spin pumping at the Py/Pd interface. A theoretical fitting was then performed in order to separate the ISHE and anomalous Hall effect (AHE) signals by using the following equation. [14]

$$V = V_{ISHE}\frac{\Gamma^2}{(H-H_{FMR})^2+\Gamma^2} + V_{AHE}\frac{-2\Gamma(H-H_{FMR})}{(H-H_{FMR})^2+\Gamma^2} + aH + b, \quad (2)$$

where $V_{ISHE}$ is the electromotive force owing to the ISHE, $V_{AHE}$ is the output voltage resulting from the AHE, and $H$ is the external static magnetic field. The value of $H_{FMR}$ was experimentally determined to be 89.51 mT at 0 and 180° at RT. The variables $\Gamma$, $a$, and $b$ are the fitting parameters. As shown in Fig. 1(e), a theoretical fitting using eq. (2)

effectively reproduced the experimental results, and $V_{ISHE}$ was estimated to be 7.66 µV at RT.

Figure 2 shows the microwave power dependence of the electromotive force from the Pd of the Py/Pd sample. The electromotive force from the Pd layer $V_{ISHE}$ was proportional to the microwave power. As shown in the inset of Fig. 2, the power dependence of $V_{ISHE}$ was in good agreement with the theoretical prediction,[2] indicating that the generation of the observed electromotive force is attributed to the ISHE in Pd. More importantly, the unsaturated FMR spectra enabled the estimation of the spin Hall angle of Pd. A pure spin current was injected at the Py/Pd interface via spin pumping under the resonance condition, and the generated spins diffused into the Pd layer and were converted to a charge current ($j_c$), as shown in Fig. 1(a). The spin current density $j_s$ was theoretically calculated as [15] $j_s = \dfrac{g_r^{\uparrow\downarrow} \gamma^2 h^2 \hbar [4\pi M_s \gamma + \sqrt{(4\pi M_s)^2 \gamma^2 + 4\omega^2}]}{8\pi\alpha^2 [(4\pi M_s)^2 \gamma^2 + 4\omega^2]}$, where $h$ is the microwave magnetic field, and it was set to 0.16 mT at a microwave power of 200 mW. The variable $g_r^{\uparrow\downarrow}$ and the constant $\hbar$ are the real part of the mixing conductance [14] and the Dirac constant, respectively. Note that $g_r^{\uparrow\downarrow}$ is given by $g_r^{\uparrow\downarrow} = \dfrac{2\sqrt{3}\pi M_s \gamma d_{Py}}{g\mu_B \omega}(W_{Py/Pd} - W_{Py})$,[7,15] where $g, \mu_B, d_{Py}, W_{Py/Pd}$, and $W_{Py}$ are the g-factor, Bohr magneton, thickness of the Py layer, FMR spectral width of the Py/Pd film, and FMR spectral width of the Py film, respectively. In this study, $d_{Py}$, $W_{Py/Pd}$, and $W_{Py}$ were 25 nm, 3.03 mT, and 2.49 mT, respectively. The electromotive force due to the ISHE can then be expressed in the simplest form as follows:[7,15]

$$V_{ISHE} = \dfrac{w \theta_{SHE} \lambda_{Pd} \tanh(d_{Pd}/2\lambda_{Pd})}{d_{Py}\sigma_{Py} + d_{Pd}\sigma_{Pd}}(\dfrac{2e}{\hbar})j_s, \qquad (3)$$

where $w$ is the length of the Py layer defined as in Fig. 1(a), $d_{Py}$ and $\sigma_{Py}$ are the thickness (25 nm) and electric conductivity ($4.48 \times 10^6$ $\Omega^{-1}$ m$^{-1}$) of the Py layer at RT, and $d_{Pd}$ and $\sigma_{Pd}$ are the thickness (5 nm) and electric conductivity ($3.50 \times 10^6$ $\Omega^{-1}$ m$^{-1}$) of the Pd layer, respectively. From the above calculations, the spin Hall angle $\theta_{SHE}$ was estimated to be 0.011 at RT, which is reasonably consistent with that obtained in a previous study.[7]

To estimate the spin Hall angle at various temperatures, the temperature evolution of the linewidth ($W$) of a simple Py film, resonance field ($H_{FMR}$), and saturation magnetization ($4\pi M_s$) were measured and evaluated as shown in Fig. 3(a), and the results were physically reasonable. Here, note that $\sigma_{Py}$, $\sigma_{Pd}$, $V_{ISHE}$, and $\lambda_{Pd}$ changed with the temperature, and thus, the temperature dependences of $\sigma_{Py}$ and $\sigma_{Pd}$ were also evaluated [Fig. 3(b)], and the estimated values of $V_{ISHE}$ obtained from eq. (2) at different temperatures are shown in the inset of Fig. 3(c). The spin diffusion length of Pd, denoted as $\lambda_{Pd}$, has been reported to be 9 nm at RT and 25 nm at 4.2 K, [16,17] so $\lambda_{Pd}$ can be estimated at each temperature by interpolating the values at RT and at 4.2 K (In fact, $\lambda_{Pd}$ is estimated to be 9.12 nm at 133 K by our calculation), since the spin lifetime is inversely proportional to temperature in this temperature region. The change in $\lambda_{Pd}$ yields little contribution to the change in $\theta_{SHE}$. Thus, $\theta_{SHE}$ of Pd can be estimated by solving eq. (3) and using the conductivities of Pd and Py and the estimated spin diffusion length of Pd at each temperature. The results of this evaluation are shown in Fig. 3(d). It was found that $\theta_{SHE}$ increased to 0.020 monotonically as the temperature decreased to 130 K. It is surprising that the temperature evolution of $\theta_{SHE}$ of Pd exhibits opposite behavior compared with that of Pt[18,19]. Whereas Pt and Pd belong to the same group in the periodic table, the electron configurations are different, which may induce such difference [20]. In fact, their electron configuration of the d-orbital allows a different sign of $\theta_{SHE}$ [21]. The origin of this characteristic temperature dependence of $\theta_{SHE}$ of Pd is beyond the scope of this study, but will be investigated in the near future.

In conclusion, in this study, the temperature dependence of the spin Hall angle of Pd was estimated by spin pumping. This approach enables the quantitative estimation of the temperature dependence of the spin transport properties of Si, graphene, and other materials by combining spin pumping and ISHE.

**Figure captions**

Figure 1(a) Schematic illustration of the Py/Pd bilayer sample used in this study. *H* represents an external magnetic field and $\theta_H$ the angle of the external magnetic field. The dimensions of the sample are shown in the figure. (b) FMR spectra of the Py layer (red line) and the Py/Pd layer on the SiO$_2$ substrate (black line). An increase in the line width, *W*, can be observed, which is attributed to the modulation of the Gilbert damping constant α, indicating successful spin injection into the Pd layer by spin pumping. (c) Magnetic field angle ($\theta_H$) dependence of the FMR signal *dI(H)/dH* for the Py/Pd bilayer sample at room temperature (RT). (d) Magnetic field angle ($\theta_H$) dependence of the electromotive force measured from the Py/Pd bilayer sample at RT. (e) External magnetic field dependence of the electromotive force $V_{ISHE}$ for the Py/Pd bilayer film at 200 mW and at RT. Open circles represent the experimental data. The solid curve shows the fitting result, which was calculated using eq. (2). Orange and blue dashed lines show the contribution from ISHE and AHE, respectively.

Figure 2. Microwave power dependence of the electromotive force in the Pd film. A magnetic field was applied parallel to the film plane ($\theta_H = 0°$). The output voltage increased with increasing microwave power. The inset shows the power dependence of the contribution of ISHE.

Figure 3(a). (upper panel) Temperature dependence of the FMR linewidth of the single Py film and the Pd/Py film. The inset shows the difference in the linewidth between the Py and Pd/Py films as a function of temperature. (lower panel) Ferromagnetic resonance field ($H_{FMR}$) and saturation magnetization ($4\pi M$s). Saturation magnetization decreased with increasing temperature. An increase in the resonance field with increasing temperature can also be observed owing to the relation shown in eq. (1). (b) Temperature dependence of the resistivity of the Pd and Py

layers. Red and blue circles indicate the experimental data for Pd and Py, respectively. Red and blue lines show the linear fittings. (c) Temperature dependence of the output electromotive force measured for the Py/Pd sample. The insets show the temperature dependence of the contribution from ISHE ($V_{\text{ISHE}}$) and an expanded graph of the electromotive force. (d) Temperature dependence of the spin Hall angle $\theta_{\text{SHE}}$ of Pd. The spin Hall angle decreased with increasing temperature.

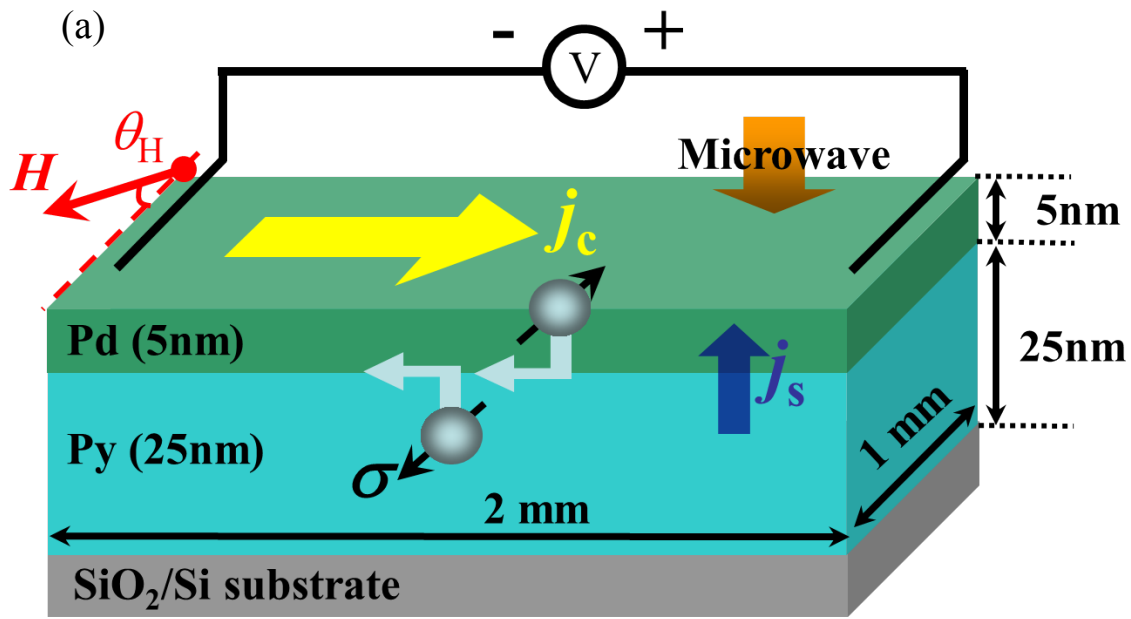

Figure 1(a).

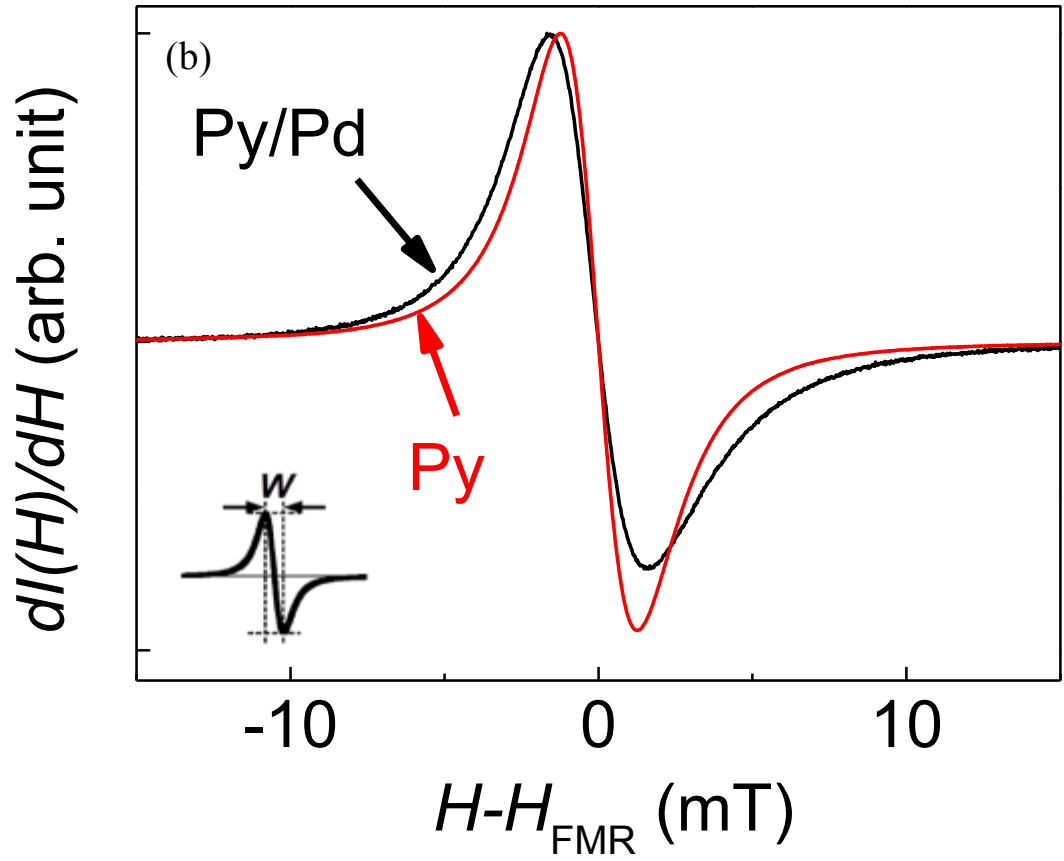

Figure 1(b).

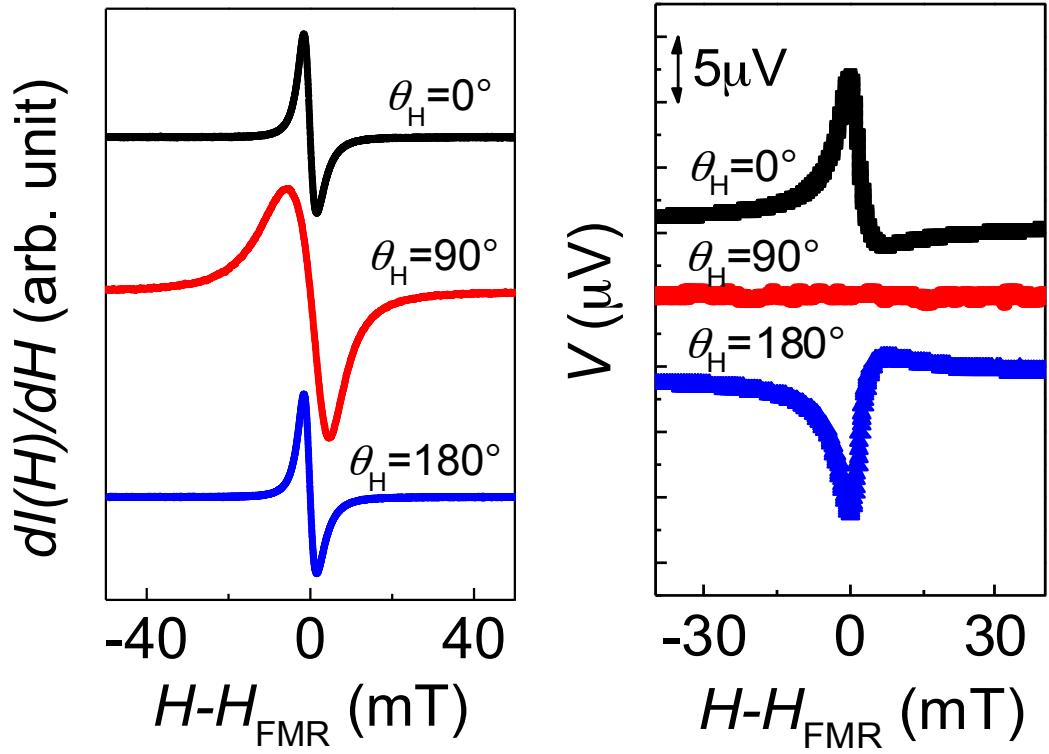

Figures 1(c) and (d)

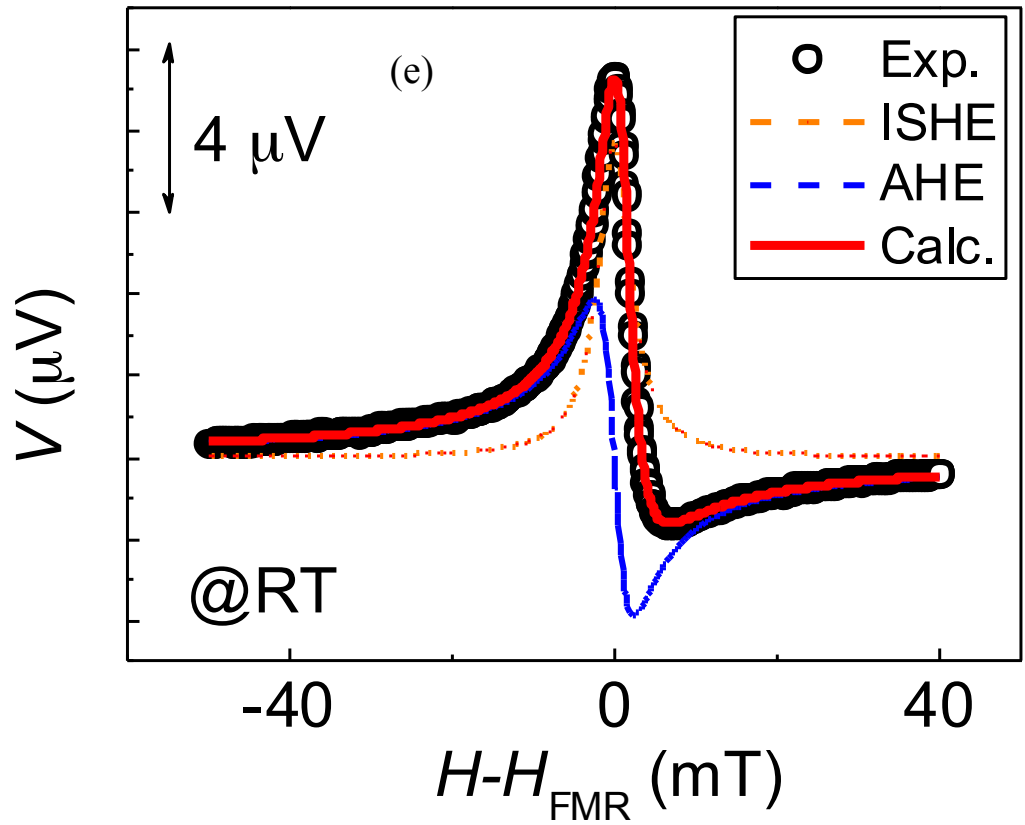

Figure 1(e)

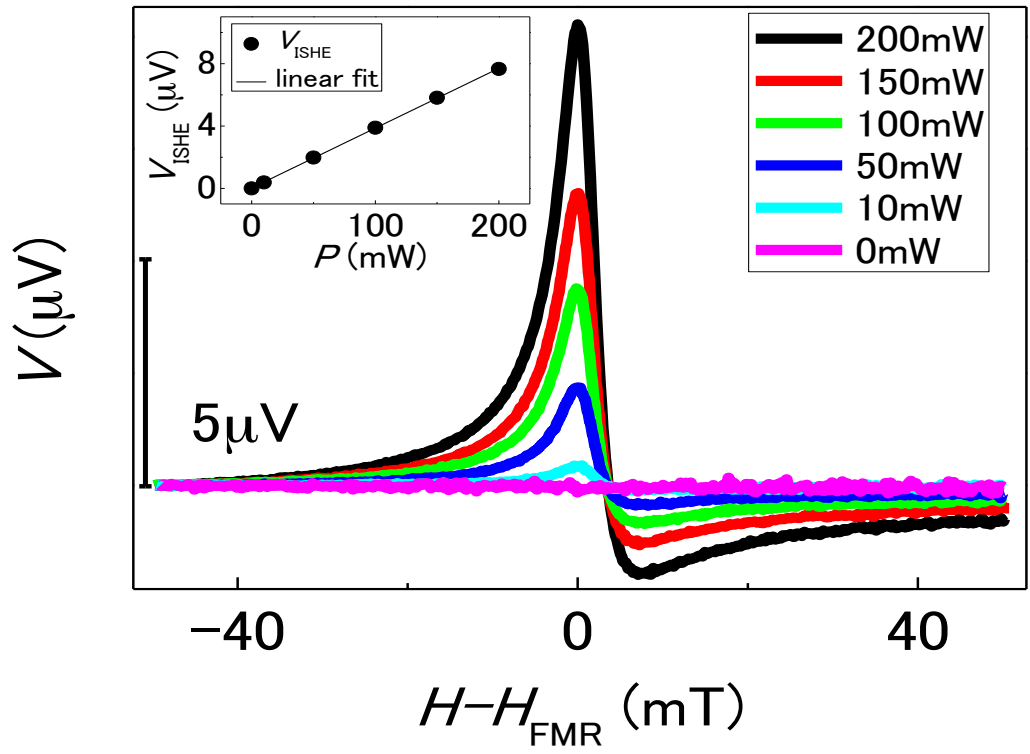

Figure 2.

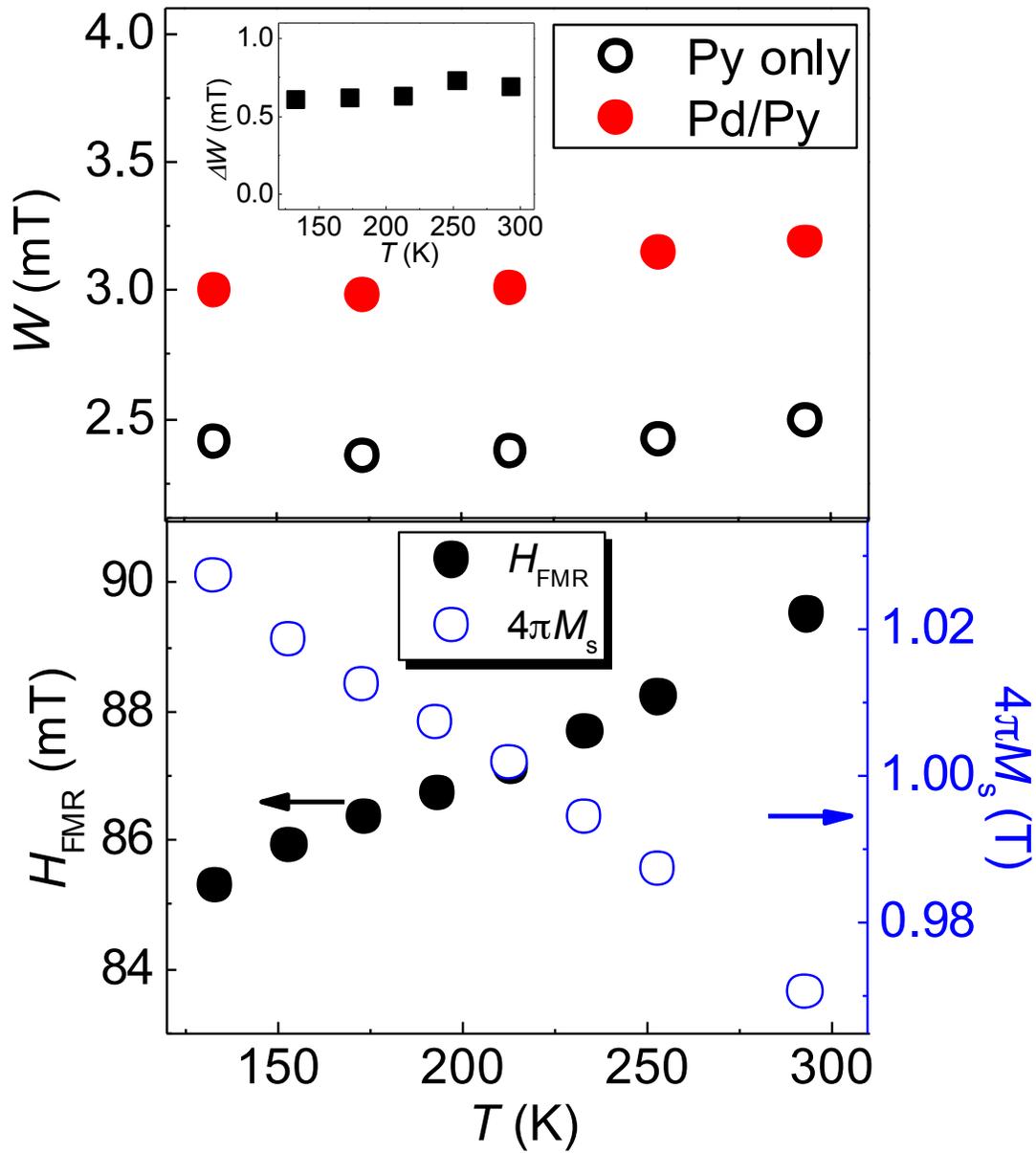

Figure 3(a).

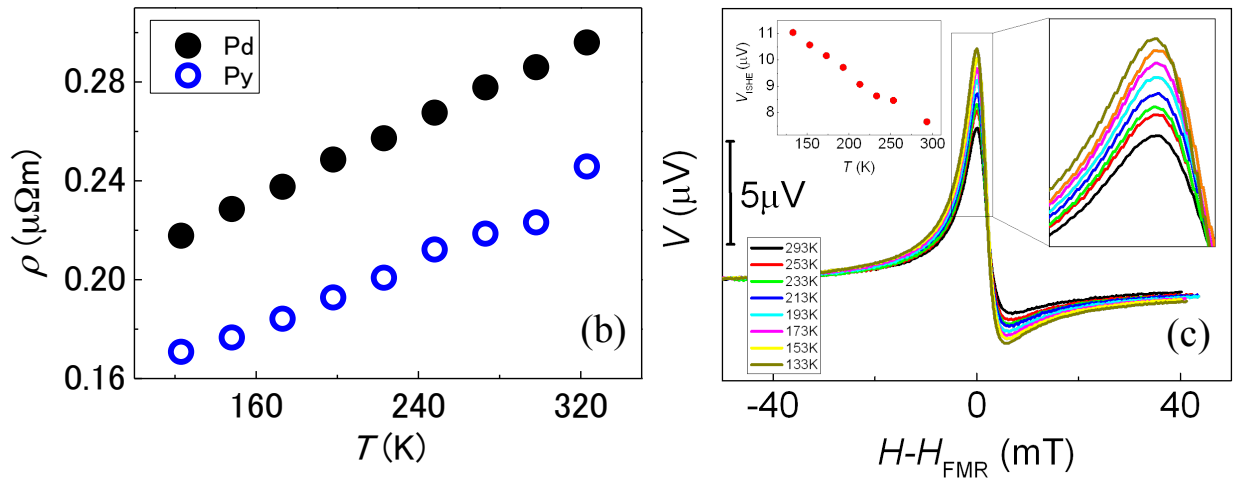

Figures 3(b) and (c)

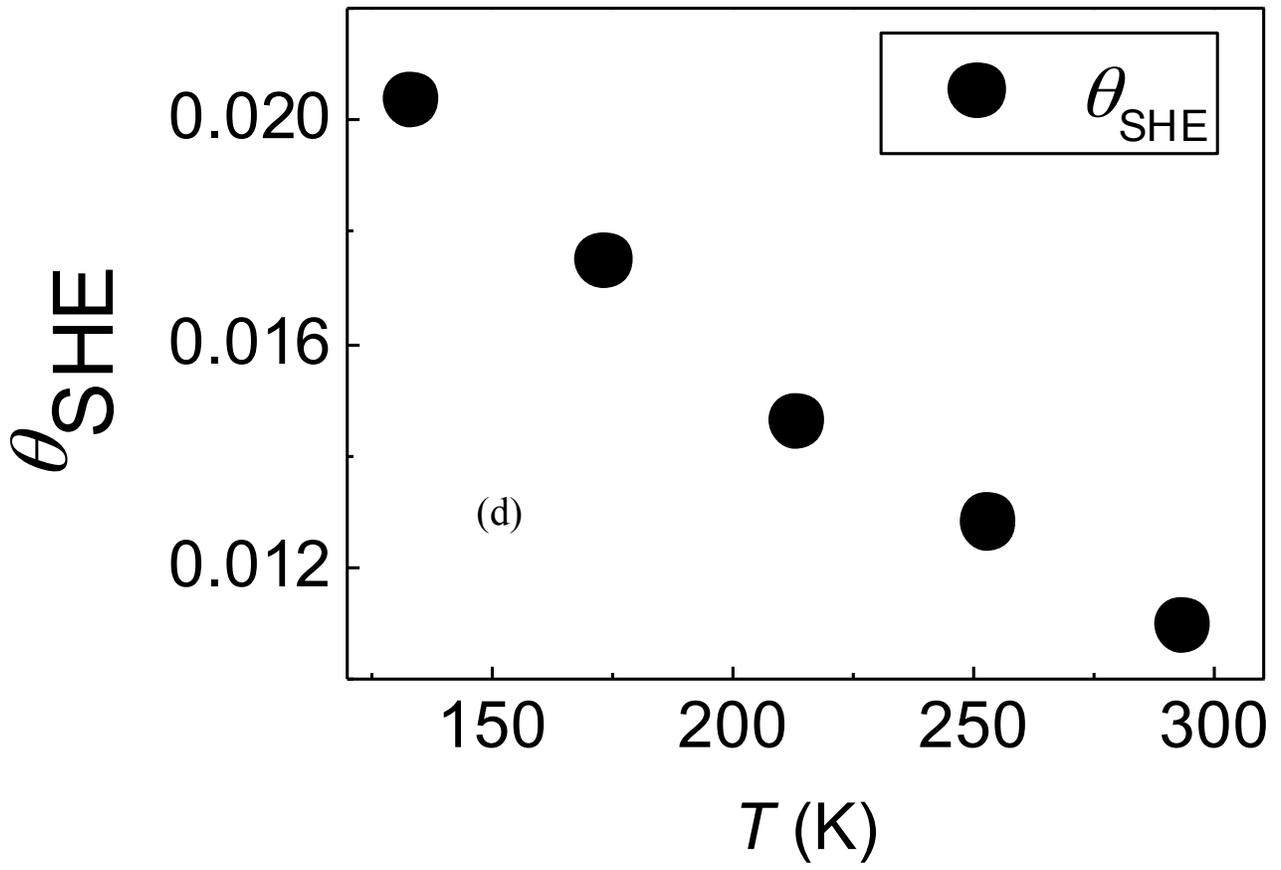

Figure 3(d)